\documentclass[pra,twocolumn,showpacs]{revtex4}
\usepackage{amsmath}
\usepackage{dcolumn}

\begin{document}
\title{Quadrupole moments and hyperfine constants of
metastable states of Ca$^+$, Sr$^+$, Ba$^+$, Yb$^+$, Hg$^+$, and Au}
\thanks{This work is a contribution of NIST, an agency of the U. S.
government, and is not subject to U. S. copyright.}
\date{February 17, 2006}
\author{Wayne M. Itano}
\email[]{itano@boulder.nist.gov} \affiliation{Time and Frequency
Division, National Institute of Standards and Technology, Boulder,
Colorado 80305}
\begin{abstract}
Atomic quadrupole moments and hyperfine constants of the metastable
$^2 D_{3/2, 5/2}$ states of Ca$^+$, Sr$^+$, Ba$^+$, Yb$^+$, and
Hg$^+$ are calculated by the multiconfiguration Dirac-Hartree-Fock
and relativistic configuration-interaction  methods. For Hg$^+$, the
configuration is $5d^9 6s^2$. For the other ions, the configuration
consists of a single $d$-electron outside a set of closed shells.
Current interest in the quadrupole moments of these states is due to
the fact that optical transitions of these ions may be useful as
references for frequency standards. Energy shifts of the metastable
states due to the interactions of the quadrupole moments with
external electric field gradients are among the largest sources of
error in these frequency standards. For the quadrupole moments,
agreement is obtained to within about 10\% with the available
measurements.  For the hyperfine constants, good agreement is
obtained with measurements and with other calculations, except for
the $A$ factors of the $^2D_{5/2}$ states of Sr$^+$, Ba$^+$, and
Yb$^+$, where the correlation effects are so large that they reverse
the sign of the constant relative to the Dirac-Hartree-Fock value.
As a test of the Hg$^+$ calculational methods, quadrupole moments
and hyperfine constants are calculated for the $5d^9 6s^2\,^2
D_{3/2, 5/2}$ states in isoelectronic neutral Au. This yields a new
value of the nuclear quadrupole moment $Q$($^{197}$Au) = +0.587(29)
b.
\end{abstract}
 \maketitle

\section{Introduction}

The electric quadrupole moment $\mathit{\Theta}(\gamma,J)$ of an
atom in  electronic state $|\gamma J\rangle$ having total electronic
angular momentum $J$ is conventionally defined by  the diagonal
matrix element in the sublevel with the maximum value of the
magnetic quantum number $M_J$:
\begin{equation}
\mathit{\Theta}(\gamma,J)=-e\left\langle \gamma J J\left|\sum_{i=1}^N r_i^2
C^{2}_0(\theta_i,\phi_i)\right|\gamma J J\right\rangle, \label{quadmoment1}
\end{equation}
where $e$ is the elementary charge, $r_i$ is the radial coordinate
of the $i$th electron, $C^{2}_0$ is a spherical harmonic, $\theta_i$
and $\phi_i$ are the angular coordinates of the $i$th electron, and
the summation is over all $N$ electrons.

In comparison with other atomic properties,  such as oscillator
strengths or hyperfine constants, atomic quadrupole moments have
received little theoretical attention, due in part to the lack of
experimental data. Only a few atomic quadrupole moments have been
measured, and most of the theoretical work appears to have been
focused on these cases.

The quadrupole moments of the  $3p\;^2P_{3/2}$ state of Al, the
$5p\;^2P_{3/2}$ state of In, and  the metastable $np^5(n+1)s\;^3P_2$
states of Ne, Ar, Kr, and Xe ($n=2,3,4,$ and 5,  respectively) were
measured by atomic beam radio frequency spectroscopy
\cite{angel67,sandars73}. In these experiments, energy level shifts
were observed upon application of an external electric field
gradient. The quadrupole moments of the 4~$^3P$ and 5~$^3P$ excited
states of He were determined indirectly from measurements of the
anisotropy of the diamagnetic susceptibility
\cite{miller71,miller72}. Since the quadrupole moment of the ionic
core influences the fine-structure of nonpenetrating Rydberg states,
it is possible to extract the quadrupole moment of the ion from an
analysis of the Rydberg spectrum of the neutral atom \cite{chang81}.
The quadrupole moments of several atomic ions, including C$^+$
\cite{chang81,ward94}, Ne$^+$
\cite{chang81,chang94,ward94,ward96,komara99}, and N$^+$
\cite{chang81,jacobson96}, have been determined in this way.

Current interest in the quadrupole moments of the metastable states
of certain atomic ions stems from the application  of narrow optical
transitions to frequency standards \cite{gill03}.  For several ions
that might be used for frequency standards, including Ca$^+$,
Sr$^+$, Ba$^+$, Yb$^+$, and Hg$^+$, the energy shifts due to the
interaction of the quadrupole moments of the metastable states with
stray electric field gradients, due, for example, to stray electric
charges on ion trap electrodes, are among the largest sources of
systematic error. This problem was first pointed out by Dehmelt
\cite{dehmelt82}.  In practice, it does not affect neutral-atom
optical frequency standards to the same extent because of the
absence of nearby charged objects. Recently, the quadrupole moments
of the $4d\; ^2D_{5/2}$ state of Sr$^+$ \cite{barwood04}, the $5d\;
^2D_{3/2}$ state of Yb$^+$ \cite{schneider05}, and the $5d^9
6s{^2}\; {^2}D_{5/2}$ state of Hg$^+$ \cite{oskay05} were determined
by observing the changes in the optical transition frequencies as
static electric field gradients were applied.

In a first approximation,  the metastable $nd\;^2 D_{3/2, 5/2}$
states of Ca$^+$, Sr$^+$, Ba$^+$, and Yb$^+$ ($n=3,4,5,$ and 5,
respectively) are described by a single configuration involving one
$d$-electron outside a set of filled shells  or, in the case of
Hg$^+$ and isoelectronic Au, a single $d$-vacancy in a set of
otherwise filled shells. In this approximation, the quadrupole
moment is due entirely to the single $d$-electron or $d$-vacancy. A
single-configuration estimate of the quadrupole moment of the $5d^9
6s{^2}\; {^2}D_{5/2}$ state of Hg$^+$ was published in
Ref.~\cite{itano00}.  However, electron correlation effects can in
some cases lead to large corrections to the single-configuration
estimates for the quadrupole moments. For example, the metastable
$^3P_2$ states of the rare gases Ne, Ar, Kr, and Xe are known to
have quadrupole moments that deviate strongly from the
single-configuration values \cite{sandars73}.  For Kr and Xe, even
the signs of the quadrupole moments differ from the
single-configuration predictions.

Sternheimer obtained good agreement with experiment for  the rare
gas quadrupole moments with a perturbative model in which the outer
$(n+1)s$ orbital is polarized by the $np$ vacancy
\cite{sternheimer73a,sternheimer73b}.  Although Sternheimer's
results were in good agreement with experiment, Sundholm and Olsen
regarded this agreement as fortuitous, particularly for Xe
\cite{sundholm93}. For Xe, they showed that the $DTQ$ electron
correlation contribution (due to double, triple, and quadruple
excitations from the $5p$ and $6s$ shells), relativistic
corrections, and excitations to virtual $f$ and $g$ orbitals all
make contributions to the quadrupole moment of about the same
magnitude as the total moment. None of these effects are included in
Sternheimer's treatment. Sundholm and Olsen applied the
multiconfiguration Hartree-Fock (MCHF) method to calculate the
quadrupole moments of the rare gas metastable atoms and some other
atoms, including Al, In, Be, Ca, C$^+$, Ne$^+$, and Ar$^+$
\cite{sundholm93,sundholm92,sundholm94,sundholm99}. Agreement with
experiment is good for the cases that have been measured. The
electric quadrupole moments of the metastable $^3P_2$ states of Mg,
Ca, and Sr have been calculated because of their importance to
ultracold collision processes
\cite{derevianko01,derevianko03,santra04}. Recently, Sur \textit{et
al.} have calculated the quadrupole moment of the $4d\;^2D_{5/2}$
state of Sr$^+$ by the relativistic coupled-cluster method
\cite{sur05}.

In this work, I apply the multiconfiguration Dirac-Hartree-Fock
(MCDHF) method, i.e., the relativistic generalization of the MCHF
method. In the final stages of the calculations,  relativistic
configuration-interaction (RCI) calculations are carried out, using
the orbitals determined by MCDHF.  This method of computing the
atomic wave function is similar to that used by Biero{\' n} and
co-workers to calculate atomic hyperfine constants \cite{bieron05}.
MCDHF appears not to have been applied previously to the calculation
of atomic quadrupole moments. It has the advantage of taking
relativity into account from the start, rather than as a correction
applied at the end of the calculation, as is done with the MCHF
method \cite{sundholm93}. This is especially important for heavy
atoms such as Yb$^+$ and Hg$^+$. Some preliminary results have been
published \cite{oskay05}.  In addition, the magnetic dipole ($A$)
and electric quadrupole ($B$) hyperfine constants are calculated and
compared with experiment, as an indication of the quality of the
wave functions. The calculation for Hg$^+$ was more complex than for
the other ions, because of the presence of the open $d$-shell. Also,
there are few measurements of the hyperfine constants of the $5d^9
6s{^2}\; {^2}D_{3/2,5/2}$ states.  For this reason, equivalent
calculations were made for Au, which is isoelectronic to Hg$^+$, and
for which the $A$ and $B$ hyperfine constants have been measured for
both fine-structure states \cite{childs66,blachman67}.

\section{Methods}

\subsection{Single-configuration estimates of the quadrupole moment}

In  a single-configuration Hartree-Fock (HF) or Dirac-Hartree-Fock
(DHF) approximation, $\mathit{\Theta}(\gamma,J)$ depends only on the
mean values of $r^2$ for the electrons which are not in closed
shells. For a configuration consisting of a single $nd$-electron
outside a set of filled shells, the electric quadrupole moments for
the $J=3/2$ and $J=5/2$ states are
\begin{subequations}
\label{singleconf1}
\begin{eqnarray}
\mathit{\Theta}(nd,3/2)&=&\frac{e}{5}\langle nd_{3/2}|r^2|
nd_{3/2}\rangle,\label{singleconf1a}\\
 \mathit{\Theta}(nd,5/2)&=&\frac{2e}{7}\langle
nd_{5/2}|r^2| nd_{5/2}\rangle.\label{singleconf1b}
\end{eqnarray}
\end{subequations}
For a $nd^9 n^{\prime}s{^2}$ configuration as in Hg$^+$,
Eqs.(\ref{singleconf1a}) and (\ref{singleconf1b}) hold with a change
of sign because the electric quadrupole moment is due to a single
vacancy in an otherwise filled shell rather than to a single
electron. In the nonrelativistic Hartree-Fock  approximation,
$\langle nd_j|r^2|nd_j\rangle$ does not depend on  $j$. Several
estimates of $\mathit{\Theta}(\gamma,J)$ based on Eqs.
(\ref{singleconf1a}) and (\ref{singleconf1b}) have appeared in the
literature \cite{itano00,barwood04,madej04,champenois04}. The radial
matrix elements were estimated from Cowan's Hartree-Fock program
\cite{cowan81} or from simple Coulombic wave functions.

\subsection{Multiconfiguration Dirac-Hartree-Fock method}

One  method of obtaining an approximation to the relativistic atomic
wave function is the MCDHF method \cite{grant94}. In the MCDHF
method an atomic state function $|\mathit{\Gamma} PJM_J\rangle$ of
parity $P$, electronic angular momentum $J$, and $z$-component of
electronic angular momentum $M_J$ is taken to be a linear
combination of relativistic configurational state functions (CSFs)
$|\gamma_k PJM_J\rangle$:
\begin{equation}
|\mathit{\Gamma} PJM_J\rangle = \sum_{k} c_k |\gamma_k
PJM_J\rangle,\label{statefunc}
\end{equation}
where each CSF is a linear combination of antisymmetrized product
wave functions (Slater determinants) such that the CSF has definite
values of $P$, $J$, and $M_J$. The CSFs differ from one another by
the orbitals (single-electron radial functions) that are occupied
and in the ways in which the angular momenta of the electrons are
coupled together.

In a MCDHF calculation, the atomic Hamiltonian is usually taken to
be the Dirac-Coulomb Hamiltonian, which includes the kinetic energy
of each electron and the Coulomb interactions of each electron with
the nucleus and with the other electrons. Additional terms, such as
the Breit interaction, may be included but increase the difficulty
of the calculation. Solving the MCDHF equations then determines an
approximate eigenfunction of the Dirac-Coulomb Hamiltonian having
the form of Eq. (\ref{statefunc}) by optimizing both the orbitals
and the coefficients $c_k$.

Once a set of orbitals has been determined by MCDHF using a limited
set of CSFs, the atomic state function can be improved by a RCI
calculation, in which the Dirac-Coulomb Hamiltonian matrix (with or
without the Breit interaction) is diagonalized in a basis consisting
of an expanded set of CSFs. The new CSFs are generated from the
orbitals calculated in the previous steps. The result of an RCI
calculation is an atomic state function having the form of Eq.
(\ref{statefunc}), but only the coefficients and not the orbitals
are optimized.  Given an approximate atomic state function, obtained
by either  MCDHF or  RCI, the atomic quadrupole moment can be
calculated by evaluating Eq. (\ref{quadmoment1}).

\subsection{Calculational details}

In the present work, the MCDHF and RCI calculations were carried out
with versions of the \textsc{GRASP} (\textbf{G}eneral-Purpose
\textbf{R}elativistic \textbf{A}tomic \textbf{S}tructure
\textbf{P}rogram) code \cite{grant80,dyall89,parpia96}. Modules from
the \textsc{GRASP92} version, documented in Ref.~\cite{parpia96},
and the \textsc{GRASPVU} version, available from a website
\cite{froesefischerwebsite} were used.

Successively improved approximations to the atomic state functions
were made in three stages. First, the orbitals belonging to the
shells that are occupied in the lowest-order approximation were
calculated by minimizing an energy functional that weighted the
$^2D_{3/2}$ and $^2D_{5/2}$ states by their statistical ($2J+1$)
weights. This is called an extended-optimal-level (EOL) calculation
\cite{dyall89}. For example, in the calculation for the Ca$^+$
$3d\;^2D_{3/2}$ and $3d\;^2D_{5/2}$ states, the $1s$, $2s$, $2p$,
$3s$, $3p$, and $3d$ orbitals were optimized.  (Here, $2p$ refers to
both the $2p_{1/2}$ and $2p_{3/2}$ orbitals, etc.) Orbitals
belonging to the same angular momentum were required to be
orthogonal. A Fermi model was used for the nuclear charge
distribution.  The Breit interaction, QED effects, and finite
nuclear mass effects were ignored throughout the calculation.

In the second stage of the calculation, several layers of virtual
orbitals were successively optimized in a series of MCDHF-EOL
calculations. Each layer consisted of a set of orbitals having
different angular momenta. All previously calculated orbitals were
kept fixed, and only the new orbitals were optimized. Different
orbitals of the same angular momentum were required to be
orthogonal. A limited set of CSFs was considered. CSFs generated by
allowing excitations of valence electrons, with or without single
excitations of certain core shells, were included.

In the final stage of the calculation, the set of CSFs was
systematically increased by allowing single excitations from
lower-lying core shells with or without valence excitations
(core-valence correlation) and double or triple excitations from
some of the higher-lying core shells (core-core correlation) to
unoccupied shells. RCI calculations were then carried out in the
expanded basis of CSFs, using the orbitals determined in the
previous stage. A practical limit to the number of CSFs in a single
RCI calculation was somewhat above 45\;000 for a single $J$-value.
The general method is similar to that used by Biero\'{n} \textit{et
al.} for the calculation of hyperfine constants of neutral mercury
\cite{bieron05}. At each step of the calculation, the hyperfine
constants and atomic quadrupole moments were calculated. The program
\textsc{HFS92} \cite{jonsson96} was used to calculate the $A$ and
$B$ hyperfine constants. I made a minor modification to the $B$
constant part of \textsc{HFS92} to enable it to calculate atomic
quadrupole moments. In some cases, core-valence contributions to the
quadrupole moment and to the hyperfine constants from different core
shells were calculated in separate RCI calculations and then
combined, making use of the fact that such contributions are
approximately additive.

Just before this paper was submitted for publication, the author
learned of an error in the \textsc{GRASP} codes, specifically in the
library function tnsrjj.f \cite{gaigalas}. The calculations were
repeated with the corrected codes. In some cases, the values of the
atomic quadrupole moments and the $B$ factors calculated with the
corrected codes differ by as much as a few percent from those
calculated with the uncorrected codes.  The values of the $A$
factors are not affected to the same extent.

\subsection{Nuclear models and moments}

For each of the atoms studied, Ca$^+$, Sr$^+$, Ba$^+$, Yb$^+$,
Hg$^+$, and Au, a particular isotope was chosen to define the
nuclear charge distribution $\rho(r)$ used for the Dirac-Coulomb
Hamiltonian used for the MCDHF and RCI calculations.  A Fermi
distribution of the form
\begin{equation}
\rho(r)=\frac{\rho_0}{1+e^{(r-c)/a}}\label{fermi}
\end{equation}
is assumed, with fitted values for the $a$ and $c$ parameters
\cite{parpia92}. The isotopes chosen for the calculations were
$^{43}$Ca, $^{87}$Sr, $^{137}$Ba, $^{171}$Yb, $^{199}$Hg, and
$^{197}$Au.  The $^{171}$Yb and $^{199}$Hg isotopes were chosen
because they are currently used in atomic frequency standards, but
since they both have nuclear spin $I=1/2$, they provide no
information about the electric quadrupole hyperfine structure.  For
that reason, the hyperfine constants were also calculated for
$^{173}$Yb$^+$ ($I=5/2$) and for $^{201}$Hg$^+$ ($I=3/2$). The
atomic wave functions calculated for $^{171}$Yb$^+$ and
$^{199}$Hg$^+$ were used.

\begin{table}[htb]
 \caption{Nuclear spins $I$ and quadrupole moments $Q$ used in the calculations
 of electric quadrupole hyperfine constants $B$. (1 b = $10^{-28}$ m$^2$.)}
 \label{qmoments}
\begin{ruledtabular}
\begin{tabular}{lcdl}
Nucleus&$I$& \multicolumn{1}{c}{$Q$(b)} & Ref.\\
 \hline
$^{43}$Ca & 7/2 & -0.0408(8) & \cite{pyykko01}\\
$^{87}$Sr & 9/2 & +0.335(20) & \cite{pyykko01}\\
$^{137}$Ba & 3/2 & +0.245(4) & \cite{pyykko01}\\
$^{173}$Yb & 5/2 & +2.80(4) & \cite{pyykko01}\\
$^{201}$Hg & 3/2 & +0.387(6) & \cite{bieron05}\\
$^{197}$Au & 3/2 & +0.547(16) & \cite{pyykko01}\\
\end{tabular}
\end{ruledtabular}
\end{table}

The nuclear magnetic moments are sufficiently well-known that their
uncertainties are likely to be much less than the errors in the
atomic calculations for $A$.  The values of the nuclear magnetic
moments were taken from the tables of Raghavan \cite{raghavan89}.
However, nuclear quadrupole moments ($Q$) are less well-known, since
they are not measured directly. Nuclear quadrupole moments derived
from interaction constants in atoms, molecules, or solids depend on
difficult calculations of the electric field gradients at the
nucleus. Values derived from muonic x-ray spectra are subject to
other systematic errors.  For example, some muonic determinations of
the nuclear quadrupole moment of $^{201}$Hg differ from each other
by more than their combined uncertainties \cite{bieron05}.  The $Q$
values used in these calculations are given in Table \ref{qmoments}.
Most of the values were taken from the compilation of Pyykk\"{o}
\cite{pyykko01}. It is a simple matter to rescale the $B$ constants
if better $Q$ values become available.

\begin{table*}[htbp]
 \caption{Quadrupole moments (in atomic units) and hyperfine constants (in MHz) for
 $^{43}$Ca$^+$ $3d\;^2D_{3/2, 5/2}$ states calculated with different approximations
 to the atomic state function.}
 \label{cacalc}
\begin{ruledtabular}
\begin{tabular}{cldddddd}
Step&Description& \multicolumn{1}{c}{$\mathit{\Theta}_{3/2}$} &
\multicolumn{1}{c}{$A_{3/2}$}
 & \multicolumn{1}{c}{$B_{3/2}$} &\multicolumn{1}{c}{$\mathit{\Theta}_{5/2}$} & \multicolumn{1}{c}{ $A_{5/2}$}
  & \multicolumn{1}{c}{$B_{5/2}$}\\
 \hline
1 & DHF & 1.461 &  -39.12 & -2.61 & 2.093 & -16.66 & -3.70 \\
2 & MCDHF (+Layer 1) & 1.307 & -46.69 & -1.83 & 1.872 & -8.90 & -2.60\\
3 & MCDHF (+Layers 1,2) & 1.266 & -49.83 & -2.75 & 1.815 & -7.89 & -3.91\\
4 & MCDHF (+Layers 1--3) & 1.228 & -51.47 & -2.72 & 1.759 & -6.35 & -3.86\\
5 & MCDHF (+Layers 1--4) & 1.139 & -52.51 & -2.99 & 1.633 & -6.74 & -4.25\\
6 & MCDHF (+Layers 1--5) & 1.136 & -52.60 & -2.94 & 1.629 & -6.28 & -4.17\\
7 & RCI (Step 6 + $\{3s,3p\}$ c-c to Layer 1) & 1.322 & -47.41 & -2.93 & 1.894 & -5.16 & -4.16\\
8 & RCI (Step 6 + $\{3s,3p\}$ c-c to Layers 1,2) & 1.338 & -47.27 & -2.94 & 1.917 & -4.84 & -4.18\\
\end{tabular}
\end{ruledtabular}
\end{table*}

\begin{table*}[htbp]
 \caption{Comparison of calculated and measured hyperfine constants (in MHz) for
 $^{43}$Ca$^+$ $3d\;^2D_{3/2, 5/2}$. Theoretical $B$ factors are scaled to the nuclear quadrupole moments listed
 in Table \ref{qmoments}.}
 \label{cacomp}
\begin{ruledtabular}
\begin{tabular}{lddddddd}
 &\multicolumn{1}{c}{Present calc.\footnote{Table \ref{cacalc}, Step 8.}} &
 \multicolumn{1}{c}{Other calc.\footnote{Relativistic many-body perturbation theory \cite{yu04}.}}
 & \multicolumn{1}{c}{Other calc.\footnote{Relativistic coupled-cluster theory \cite{sahoo03}.}}
  &\multicolumn{1}{c}{Other calc.\footnote{Many-body perturbation theory with relativistic correction \cite{martensson92}.}}
  &\multicolumn{1}{c}{Other calc.\footnote{Many-body perturbation theory \cite{martensson84}.}}
  & \multicolumn{1}{c}{Expt.\footnote{Reference \cite{nortershauser98}.}}
  & \multicolumn{1}{c}{Expt.\footnote{Reference \cite{kurth95}.}} \\
 \hline
 $A_{3/2}$ & -47.27 & -47.824 & -46.70  & -49.4 & -52 &  -47.3(0.2) & -48.3(1.6) \\
 $B_{3/2}$ & -2.94 & -2.777 &   &  & -2.77 & -3.7(1.9) & -0.5(6.0) \\
 $A_{5/2}$ & -4.84 & -3.552 & -3.49  & -4.2 & -5.2 &  -3.8(0.6) &  \\
 $B_{5/2}$ & -4.18 & -4.088 &   &  & -3.97 & -3.9(6.0) & \\
\end{tabular}
\end{ruledtabular}
\end{table*}

\section{Results}

\subsection{Ca$^+$}

The results of the calculation for $^{43}$Ca$^+$ are given in Table
\ref{cacalc}. DHF refers to a Dirac-Hartree-Fock EOL calculation.
Five layers of virtual orbitals were optimized in a series of
MCDHF-EOL calculations.  All CSFs having the proper parity and total
angular momentum that could be constructed by allowing single and
double excitations from the valence $3d$ and the $\{2s,2p,3s,3p\}$
core shells, with at most one core excitation, were included. The
orbitals in Layers 1 to 5 were  $\{4s,4p,4d,4f,5g,6h \}$,
$\{5s,5p,5d,5f,6g\}$, $\{6s,6p,6d,6f\}$, $\{7s,7p,7d\}$, and
$\{8s,8p,8d\}$, respectively. A limited amount of core-core (c-c)
correlation was then included by considering the CSFs obtained by
allowing double excitations from the $\{3s,3p\}$ core shells to
Layer 1 (Step 7) and to Layers 1 and 2 (Step 8).  This set of CSFs
was added to the set used in the Step 7 MCDF calculation. The atomic
state functions were then optimized in RCI calculations.  It is of
interest to note that the final values of the quadrupole moments and
the hyperfine constants are not too different from the DHF values,
except for $A_{5/2}$, which is smaller in magnitude by a factor of
3.4.

Table \ref{cacomp} compares the results of the final RCI calculation
with experiment and with other calculations. The agreement of the
calculated value for $A_{3/2}$ with both experiment and with other
recent calculations is excellent. For $A_{5/2}$, where the
corrections to the DHF value are very large, the calculated value
disagrees with experiment by about two experimental standard
deviations. The calculated $B_{3/2}$ and $B_{5/2}$ factors are in
agreement with experiment, but the experimental uncertainties are
large. They agree well with other calculations.

\begin{table*}[htbp]
 \caption{Quadrupole moments (in atomic units) and hyperfine constants (in MHz) for
 $^{87}$Sr$^+$ $4d\;^2D_{3/2, 5/2}$ states calculated with different approximations to the atomic state function.}
 \label{srcalc}
\begin{ruledtabular}
\begin{tabular}{cldddddd}
Step&Description& \multicolumn{1}{c}{$\mathit{\Theta}_{3/2}$} &
\multicolumn{1}{c}{$A_{3/2}$}
 & \multicolumn{1}{c}{$B_{3/2}$} &\multicolumn{1}{c}{$\mathit{\Theta}_{5/2}$} & \multicolumn{1}{c}{ $A_{5/2}$}
  & \multicolumn{1}{c}{$B_{5/2}$}\\
 \hline
1 & DHF & 2.309 &  -34.23 & 29.56 & 3.332 & -14.27 & 40.45 \\
2 & MCDHF (+Layer 1) & 2.083 & -45.14 & 23.47 & 3.009 & 0.57 & 32.34\\
3 & MCDHF (+Layers 1,2) & 2.021 & -47.32 & 32.66 & 2.921 & -1.01 & 45.33\\
4 & MCDHF (+Layers 1--3) & 1.966 & -50.96 & 33.73 & 2.843 & -1.38 & 46.68\\
5 & MCDHF (+Layers 1--4) & 1.847 & -51.36 & 35.10 & 2.674 & -2.31 & 48.63\\
6 & MCDHF (+Layers 1--5) & 1.844 & -51.59 & 34.76 & 2.670 & -2.07 & 48.14\\
7 & RCI (Step 6 + $\{3s,3p\}$ c-v) & 1.831 & -52.20 & 41.92 & 2.651 & -2.64 & 58.14\\
8 & RCI (Step 6 + $\{3d,4s,4p\}$ c-c to Layer 1) & 2.093 & -46.13 & 34.96 & 3.028 & -2.61 & 48.38\\
9 & RCI (Step 6 + $\{4s,4p\}$ c-c to Layers 1,2) & 2.101 & -45.14 & 34.04 & 3.038 & -2.12 & 47.13\\
10 & RCI (Step 8 $\bigcup$ Step 9)  & 2.117 & -45.26 & 34.30 & 3.061 & -2.37 & 47.50\\
11 & RCI (Step 7 $\bigcup$ Step 8 $\bigcup$ Step 9)  & 2.107 & -45.60 & 41.04 & 3.048 & -2.77 & 56.94\\
\end{tabular}
\end{ruledtabular}
\end{table*}

\begin{table*}[htbp]
 \caption{Comparison of calculated and measured quadrupole moments (in atomic units) and hyperfine constants
 (in MHz) for
 $^{87}$Sr$^+$ $4d\;^2D_{3/2, 5/2}$. Theoretical $B$ factors are scaled to the nuclear quadrupole moments listed
 in Table \ref{qmoments}.}
 \label{srcomp}
\begin{ruledtabular}
\begin{tabular}{ldddddl}
 &\multicolumn{1}{c}{Present calc.\footnote{Table \ref{srcalc}, Step 11.}} &
 \multicolumn{1}{c}{Other calc.\footnote{Relativistic many-body perturbation theory \cite{yu04}.}}
 & \multicolumn{1}{c}{Other calc.\footnote{Relativistic coupled-cluster theory \cite{martensson02}.}}
&\multicolumn{1}{c}{Other calc.\footnote{Relativistic
coupled-cluster theory \cite{sur05}.}}
  & \multicolumn{1}{c}{Expt.\footnote{Reference \cite{barwood03}.}}
  & \multicolumn{1}{l}{Expt.\footnote{Reference \cite{barwood04}.}} \\
 \hline
 $A_{3/2}$ & -45.60 & -47.356 & -47  &  &  & \\
 $B_{3/2}$ & 41.04 & 39.610 & 38.5  &  &  & \\
 $\mathit{\Theta}_{5/2}$ & 3.048 & &  & 2.94(7) & & 2.6(3)\\
 $A_{5/2}$ & -2.77 & 2.507 & 1   & & 2.1743(14) & \\
 $B_{5/2}$ & 56.94 & 56.451 & 56.0 &   & 49.11(6) & \\
\end{tabular}
\end{ruledtabular}
\end{table*}

\begin{table*}[htbp]
 \caption{Quadrupole moments (in atomic units) and hyperfine constants (in MHz) for
 $^{137}$Ba$^+$ $5d\;^2D_{3/2, 5/2}$ states calculated with different approximations to the atomic state function.}
 \label{bacalc}
\begin{ruledtabular}
\begin{tabular}{cldddddd}
Step&Description& \multicolumn{1}{c}{$\mathit{\Theta}_{3/2}$} &
\multicolumn{1}{c}{$A_{3/2}$}
 & \multicolumn{1}{c}{$B_{3/2}$} &\multicolumn{1}{c}{$\mathit{\Theta}_{5/2}$} & \multicolumn{1}{c}{ $A_{5/2}$}
  & \multicolumn{1}{c}{$B_{5/2}$}\\
 \hline
1 & DHF & 2.589 &  139.23 & 35.44 & 3.788 & 55.82 & 45.36 \\
2 & MCDHF (+Layer 1) & 2.284 & 184.37 & 24.60 & 3.354 & -20.04 & 31.98\\
3 & MCDHF (+Layers 1,2) & 2.202 & 193.43 & 35.99 & 3.239 & -10.71 & 47.79\\
4 & MCDHF (+Layers 1--3) & 2.110 & 202.28 & 35.46 & 3.116 & -9.26 & 46.82\\
5 & MCDHF (+Layers 1--4) & 2.094 & 201.80 & 36.63 & 3.088 & -6.07 & 48.72\\
6 & MCDHF (+Layers 1--5) & 2.073 & 203.12 & 36.35 & 3.051 & -7.21 & 48.01\\
7 & RCI (Step 6 + $\{4s,4p\}$ c-v) & 2.061 & 208.36 & 46.60 & 3.034 & -4.72 & 61.87\\
8 & RCI (Step 6 + $\{3s,3p,3d,4s,4p\}$ c-v) & 2.058 & 212.70 & 49.63 & 3.029 & 0.09 & 65.92\\
9 & RCI (Step 6 + $\{5s,5p\}$ c-c to Layer 1) & 2.268 & 185.86 & 39.33 & 3.338 & -2.10 & 51.98\\
10 & RCI (Step 6 + $\{5s,5p\}$ c-c to Layers 1,2) & 2.279 & 184.05 & 38.62 & 3.354 & -3.27 & 51.07\\
11 & RCI (Step 6 + $\{4d,5s,5p\}$ c-c to Layer 1) & 2.299 & 185.90 & 40.03 & 3.382 & 0.05 & 52.92\\
12 & RCI (Step 10 $\bigcup$ Step 11)  & 2.309 & 184.31 & 39.38 & 3.397 & -1.01 & 52.08\\
13 & RCI (Step 7 $\bigcup$ Step 10 $\bigcup$ Step 11)  & 2.299 & 188.65 & 48.29 & 3.384 & 4.59 & 64.11\\
14 & Step 13 + $\{3s,3p,3d\}$ c-v  & 2.297 & 192.99 & 51.32 & 3.379 & 9.39 & 68.16\\
\end{tabular}
\end{ruledtabular}
\end{table*}

\begingroup
\squeezetable
\begin{table}[htbp]
 \caption{Comparison of calculated and measured  hyperfine constants
 (in MHz) for
 $^{137}$Ba$^+$ $5d\;^2D_{3/2, 5/2}$. Theoretical $B$ factors are scaled to the nuclear
 quadrupole moments listed
 in Table \ref{qmoments}.}
 \label{bacomp}
\begin{ruledtabular}
\begin{tabular}{ldddd}
 &\multicolumn{1}{c}{Present calc.\footnote{Table \ref{bacalc}, Step 14.}}
 & \multicolumn{1}{c}{Other calc.\footnote{Relativistic coupled-cluster theory \cite{sahoo03a}.}}
  &\multicolumn{1}{c}{Other calc.\footnote{Many-body perturbation theory \cite{silverans86}.}}
  & \multicolumn{1}{c}{Expt.\footnote{Reference \cite{silverans86}.}}\\
 \hline
 $A_{3/2}$ & 192.99 & 188.76 &  215  & 189.7296(7)    \\
 $B_{3/2}$ & 51.32 &  & 47.3 &  44.5408(17) \\
 $A_{5/2}$ & 9.39 & & -18  & -12.028(11)  \\
 $B_{5/2}$ & 68.16  &   & 63.2 & 59.533(43)\\
\end{tabular}
\end{ruledtabular}
\end{table}
\endgroup

\begingroup
\squeezetable
\begin{table*}[htbp]
 \caption{Quadrupole moments (in atomic units) and hyperfine constants (in MHz) for
 $^{171}$Yb$^+$ and $^{173}$Yb$^+$ $5d\;^2D_{3/2, 5/2}$ states calculated with different approximations \
 to the atomic state function.}
 \label{ybcalc}
\begin{ruledtabular}
\begin{tabular}{clddddcdddd}
Step&Description & \multicolumn{4}{c}{$^{171}$Yb$^+$} & & \multicolumn{4}{c}{$^{173}$Yb$^+$}\\
 \cline{3-6} \cline{8-11}
 & & \multicolumn{1}{r}{$\mathit{\Theta}_{3/2}$} & \multicolumn{1}{c}{$A_{3/2}$}
 &\multicolumn{1}{r}{$\mathit{\Theta}_{5/2}$} & \multicolumn{1}{c}{ $A_{5/2}$}
 & & \multicolumn{1}{c}{$A_{3/2}$}& \multicolumn{1}{l}{$B_{3/2}$}
 & \multicolumn{1}{c}{$A_{5/2}$}& \multicolumn{1}{l}{$B_{5/2}$}
 \\
 \hline
1 & DHF & 2.343 &  319.59  & 3.467 & 121.39  & & -88.03 & 612.0 & -33.44 & 719.5 \\
2 & MCDHF (+Layer 1) & 2.157 &  383.38  & 3.204 & -82.91 & & -105.60 & 474.8 & 22.84 & 564.0 \\
3 & MCDHF (+Layers 1,2) & 1.989 &  404.88  & 2.989 & -82.82 & & -111.52 & 649.7 & 22.81 & 815.3 \\
4 & MCDHF (+Layers 1--3) & 1.891 &  437.03  & 2.847 & -87.12 & & -120.38 & 686.8 & 24.00 & 845.8 \\
5 & MCDHF (+Layers 1--4) & 1.845 &  430.20  & 2.780 & -83.48 & & -118.50 & 687.4 & 22.99 & 853.1 \\
6 & MCDHF (+Layers 1--5) & 1.823 &  435.77  & 2.743 & -83.95 & & -120.03 & 694.9 & 23.12 & 861.4 \\
7 & RCI (Step 6 + $\{4s,4p,4d\}$ c-v) & 1.805 &  467.15  & 2.718 & -61.08 & & -128.67 & 965.0 & 16.82 & 1207.3 \\
8 & RCI (Step 6 + $\{3s,3p,3d,4s,4p,4d\}$ c-v) & 1.804 &  469.95  & 2.716 & -61.07 & & -129.45 & 981.7 & 16.82 & 1228.7 \\
9 & RCI (Step 6 + $4f$ c-c to Layer 1$^\prime$) & 2.036 &  401.79  & 3.047 & -53.12 & & -110.67 & 694.0 & 14.63 & 860.5 \\
10 & RCI (Step 6 + $\{4f,5s,5p\}$ c-c to Layer 1$^\prime$) & 2.183 &  384.04  & 3.256 & -26.55 & & -105.78 & 748.8 & 7.31 & 929.8\\
11 & RCI (Step 7 $\bigcup$ Step 10) & 2.175 &  397.68  & 3.245 & -12.59 & & -109.54 & 934.7 & 3.47 & 1169.0\\
12 & Step 11 +$\{3s,3p,3d\}$ c-v & 2.174 &  400.48  & 3.244 & -12.58 & & -110.31 & 951.4 & 3.47 & 1190.4\\
\end{tabular}
\end{ruledtabular}
\end{table*}
\endgroup

\begin{table}[htbp]
 \caption{Comparison of calculated and measured quadrupole moments (in atomic units) and hyperfine constants
 (in MHz) for $^{171}$Yb$^+$ $5d\;^2D_{3/2, 5/2}$. }
 \label{ybcomp}
\begin{ruledtabular}
\begin{tabular}{ldddd}
 &\multicolumn{1}{c}{Present calc.\footnote{Table \ref{ybcalc}, Step 12.}}

  &\multicolumn{1}{c}{Expt.\footnote{Reference \cite{engelke96}.}}
  &\multicolumn{1}{c}{Expt.\footnote{Reference \cite{schneider05}.}}
  & \multicolumn{1}{c}{Expt.\footnote{Reference \cite{roberts99}.}}\\
 \hline
 $\mathit{\Theta}_{3/2}$ & 2.174  &     &  2.08(11)& \\
 $A_{3/2}$               & 400.48 & 430(43) &          &     \\
 $A_{5/2}$               & -12.58 &     &          &  -63.6(7) \\
\end{tabular}
\end{ruledtabular}
\end{table}

\begingroup
\squeezetable
\begin{table*}[htbp]
 \caption{Quadrupole moments (in atomic units) and hyperfine constants (in MHz) for
 $^{199}$Hg$^+$ and $^{201}$Hg$^+$ $5d^9 6s^2\;^2D_{3/2, 5/2}$ states calculated with different approximations \
 to the atomic state function.}
 \label{hgcalc}
\begin{ruledtabular}
\begin{tabular}{clldldcdddd}
Step&Description & \multicolumn{4}{c}{$^{199}$Hg$^+$}& & \multicolumn{4}{c}{$^{201}$Hg$^+$}\\
\cline{3-6} \cline{8-11}
 & & \multicolumn{1}{l}{$\mathit{\Theta}_{3/2}$} & \multicolumn{1}{l}{$A_{3/2}$}
 &\multicolumn{1}{l}{$\mathit{\Theta}_{5/2}$} & \multicolumn{1}{l}{ $A_{5/2}$}
 & & \multicolumn{1}{l}{$A_{3/2}$}& \multicolumn{1}{l}{$B_{3/2}$}
 & \multicolumn{1}{l}{$A_{5/2}$}& \multicolumn{1}{l}{$B_{5/2}$}
 \\
 \hline
1 & DHF                                        & -0.44575 &  2689.4  & -0.68869 &  991.1& & -992.8 & -772.2 & -365.9 & -798.8 \\
2 & MCDHF (+Layer 1)                           & -0.33741 &  2496.7  & -0.53697 & 1087.1& & -921.6 & -655.5 & -401.3 & -714.7 \\
3 & MCDHF (+Layers 1,2)                        & -0.35076 &  2507.5  & -0.55342 &  940.2& & -925.6 & -652.2 & -347.1 & -722.6 \\
4 & MCDHF (+Layers 1--3)                       & -0.34786 &  2478.6  & -0.54965 &  911.0& & -914.9 & -662.6 & -336.3 & -741.3 \\
5 & MCDHF (+Layers 1--4)                       & -0.34729 &  2457.7  & -0.54914 &  900.7& & -907.2 & -666.4 & -332.5 & -747.7 \\
6 & RCI (Step 4 + $4f$ c-v)                    & -0.34726 &  2447.8  & -0.54913 &  898.4& & -903.6 & -663.1 & -331.6 & -742.4 \\
7 & RCI (Step 4 + $4d$ c-v)                    & -0.34716 &  2525.2  & -0.54898 &  928.7& & -932.2 & -682.3 & -342.8 & -765.2 \\
8 & RCI (Step 4 + $4p$ c-v)                    & -0.34722 &  2504.4  & -0.54905 &  922.1& & -924.5 & -712.0 & -340.4 & -807.5 \\
9 & RCI (Step 4 + $4s$ c-v)                    & -0.34739 &  2452.5  & -0.54928 &  930.6& & -905.3 & -662.2 & -343.5 & -740.4 \\
10 & RCI (Step 4 + $3d$ c-v)                   & -0.34739 &  2482.8  & -0.54929 &  912.5& & -916.5 & -664.2 & -336.8 & -742.8 \\
11 & RCI (Step 4 + $\{5d,6s\}$ $d$ c-c)        & -0.36528 &  2506.3  & -0.57286 &  879.3& & -925.2 & -663.8 & -324.6 & -746.2 \\
12 & RCI (Step 4 + $\{5d,6s\}$ $dt$ c-c)       & -0.35470 &  2441.4  & -0.55744 &  962.1& & -901.2 & -661.5 & -355.1 & -742.8 \\
13 & RCI (Step 12 $\bigcup$ $\{5s,5p,5d\}$ c-c)& -0.36070 &  2458.5  & -0.56627 &  926.2& & -907.5 & -666.1 & -341.9 & -747.4 \\
14 & RCI (Step 13 + $\{3d,4spdf\}$ c-v)        & -0.35795 &  2478.3  & -0.56374 &  963.5& & -914.8 & -737.0 & -355.7 & -839.4 \\
\end{tabular}
\end{ruledtabular}
\end{table*}
\endgroup

\begingroup
\squeezetable
\begin{table}[htbp]
 \caption{Comparison of calculated and measured quadrupole moments (in atomic units) and hyperfine constants
 (in MHz) for $^{199}$Hg$^+$ and $^{201}$Hg$^+$ $5d^9 6s^2\;^2D_{3/2, 5/2}$.
  Theoretical $B$ factors are scaled to the nuclear quadrupole moments listed
 in Table \ref{qmoments}.
 }
 \label{hgcomp}
\begin{ruledtabular}
\begin{tabular}{lldddd}
 & &\multicolumn{1}{c}{Present calc.\footnote{Table \ref{hgcalc}, Step 14.}}
  &\multicolumn{1}{c}{Other calc.\footnote{MCDHF \cite{brage99}.}}
  &\multicolumn{1}{c}{Expt.\footnote{Reference \cite{itano00}.}}
  & \multicolumn{1}{c}{Expt.\footnote{Reference \cite{oskay05}.}}\\
 \hline
                &$A_{3/2}$ & 2478.3 & 2399 &           &     \\
 $^{199}$Hg$^+$&$\mathit{\Theta}_{5/2}$ & -0.56374&      &           & -0.510(18)\\
 & $A_{5/2}$ & 963.5  & 1315 &  986.19(4)&     \\
 \hline
               & $A_{3/2}$ & -914.8 & -879 &           &     \\
$^{201}$Hg$^+$ & $B_{3/2}$ & -737.0 & -674 &           &     \\
               & $A_{5/2}$ & -355.7 & -482 &           &  \\
               & $B_{5/2}$ & -839.4 & -731 &           &     \\
\end{tabular}
\end{ruledtabular}
\end{table}
\endgroup

\begingroup
\squeezetable
\begin{table*}[htbp]
 \caption{Quadrupole moments (in atomic units) and hyperfine constants (in MHz) for
 $^{197}$Au $5d^9 6s^2\;^2D_{3/2, 5/2}$ states calculated with different approximations \
 to the atomic state function.}
 \label{aucalc}
\begin{ruledtabular}
\begin{tabular}{cldddddd}
Step&Description & \multicolumn{1}{c}{$\mathit{\Theta}_{3/2}$} &
\multicolumn{1}{l}{$A_{3/2}$} & \multicolumn{1}{l}{$B_{3/2}$}
 &\multicolumn{1}{c}{$\mathit{\Theta}_{5/2}$} & \multicolumn{1}{l}{ $A_{5/2}$} & \multicolumn{1}{l}{$B_{5/2}$}\\
 \hline
1 & DHF                                        & -0.52677 &  217.34  & -855.4  & -0.82660 &  79.34  & -940.3 \\
2 & MCDHF (+Layer 1)                           & -0.40166 &  196.90  & -764.3  & -0.65285 &  91.57  & -824.8 \\
3 & MCDHF (+Layers 1,2)                        & -0.42674 &  198.07  & -756.7  & -0.68586 &  77.07  & -832.4 \\
4 & MCDHF (+Layers 1--3)                       & -0.42469 &  196.81  & -771.6  & -0.68413 &  72.44  & -859.5 \\
5 & MCDHF (+Layers 1--4)                       & -0.42348 &  195.07  & -778.2  & -0.68308 &  71.13  & -870.0 \\
6 & RCI (Step 4 + $4f$ c-v)                    & -0.42353 &  194.01  & -771.7  & -0.68316 &  71.28  & -860.2 \\
7 & RCI (Step 4 + $4d$ c-v)                    & -0.42341 &  200.50  & -795.0  & -0.68299 &  73.80  & -887.6 \\
8 & RCI (Step 4 + $4p$ c-v)                    & -0.42359 &  199.48  & -827.3  & -0.68322 &  73.12  & -932.3 \\
9 & RCI (Step 4 + $4s$ c-v)                    & -0.42364 &  194.05  & -770.8  & -0.68333 &  74.77  & -858.0 \\
10 & RCI (Step 4 + $3d$ c-v)                   & -0.42365 &  197.13  & -772.9  & -0.68331 &  72.53  & -860.6 \\
11 & RCI (Step 4 + $\{5d,6s\}$ $d$ c-c)        & -0.44945 &  200.65  & -774.8  & -0.71684 &  68.32  & -868.8 \\
12 & RCI (Step 4 + $\{5d,6s\}$ $dt$ c-c)       & -0.43511 &  192.36  & -770.2  & -0.69560 &  78.59  & -861.6 \\
13 & RCI (Step 12 $\bigcup$ $\{5s,5p,5d\}$ c-c)& -0.44396 &  194.85  & -777.0  & -0.70940 &  74.50  & -868.9 \\
14 & RCI (Step 13 + $\{3d,4spdf\}$ c-v)        & -0.43831 &  195.96  & -856.9  & -0.70473 &  77.77  & -970.2 \\
\end{tabular}
\end{ruledtabular}
\end{table*}
\endgroup

\begin{table}
 \caption{Comparison of calculated and measured  hyperfine constants
 (in MHz) for $^{197}$Au $5d^9 6s^2\;^2D_{3/2, 5/2}$. Theoretical $B$ factors are scaled to the nuclear
 quadrupole moments listed in Table \ref{qmoments}.}
 \label{aucomp}
\begin{ruledtabular}
\begin{tabular}{lddd}
 &\multicolumn{1}{c}{Present calc.\footnote{Table \ref{aucalc}, Step 14.}}
  &\multicolumn{1}{c}{Expt.\footnote{Reference \cite{blachman67}.}}
  &\multicolumn{1}{c}{Expt.\footnote{Reference \cite{childs66}.}}
  \\
 \hline

 $A_{3/2}$  & 195.96 & 199.8425(2)     &          \\
 $B_{3/2}$  & -856.9 & -911.0766(5)    &          \\
 $A_{5/2}$  &  77.77 &                 & 80.236(3)\\
 $B_{5/2}$  & -970.2 &                 & -1049.781(11) \\
\end{tabular}
\end{ruledtabular}
\end{table}

\subsection{Sr$^+$}

The results of the calculation for $^{87}$Sr$^+$ are given in Table
\ref{srcalc}. Similarly to the calculation for Ca$^+$, five layers
of virtual orbitals were optimized in a series of MCDHF-EOL
calculations. Single and double excitations from the valence $4d$
and the $\{3d,4s,4p\}$ core shells, with at most one core
excitation, were included. The orbitals in Layers 1 to 5 were
$\{5s,5p,5d,4f,5g,6h \}$, $\{6s,6p,6d,5f,6g\}$, $\{7s,7p,7d,6f\}$,
$\{8s,8p,8d\}$, and $\{9s,9p,9d\}$, respectively. Step 7 is an  RCI
calculation including the CSFs of Step 6 and also those involving
single excitations from the $\{3s,3p\}$ shells. This adds
core-valence correlation not already included in the MCDHF
calculations. Other RCI calculations including core-core correlation
were made, allowing double excitations from the $\{3d,4s,4p\}$
shells to Layer 1 (Step 8) and allowing double excitations from the
$\{4s,4p\}$ shells to Layers 1 and 2 (Step 9), in addition to the
Step 6 CSFs. Step 10 is an RCI calculation with a set of CSFs that
is the union of the sets used for Steps 8 and 9. Step 11 uses the
union of the sets of CSFs used for Steps 7 and 10.

Table \ref{srcomp} compares the final results with experiment and
with other calculations. There are no experimental results for the
$4d\;^2D_{3/2}$ state.  The present results are generally in good
agreement with other calculations and with the available
experimental data, with the exception of $A_{5/2}$.  It appears that
the correlation correction to $A_{5/2}$ is so large as to change its
sign relative to the DHF value.  The difference between the present
calculation and the experimental value is -4.94 MHz, which is 30\%
of the total correlation correction, obtained by taking the
difference of  the DHF value and the experimental value. The fact
that three calculations yield a value for $B_{5/2}$ of approximately
56 MHz, while the experimental value is 49.11(6) MHz, suggests that
the $^{87}$Sr nuclear quadrupole might be approximately 0.29 b,
rather than 0.335(20) b as listed in Table \ref{qmoments}.  The
value in Table \ref{qmoments} is based on the measured $B$ factor of
the $5s5p\;^3P_2$ state of $^{87}$Sr, combined with a semiempirical
calculation of the electric field gradient based on the measured
magnetic dipole hyperfine interaction constants \cite{heider77}.  A
calculated Sternheimer antishielding correction \cite{sternheimer50}
was applied.  Some other recent determinations of $Q$ for $^{87}$Sr,
based on the measured $B$ factor of the $5p\;^2P_{3/2}$ state of
$^{87}$Sr$^+$ combined with \textit{ab initio} calculations of the
electric field gradient, are 0.327(24) b \cite{martensson02} and
0.323(20) b \cite{yu04}. The present calculation of
$\mathit{\Theta}_{5/2}$ agrees to within 4\% with a recent
relativistic coupled-cluster calculation.  The measured value of
$\mathit{\Theta}_{5/2}$ disagrees with the present calculation by
about 1.5 standard deviations of the measurement.

\subsection{Ba$^+$}

The results of the calculation for $^{137}$Ba$^+$ are given in Table
\ref{bacalc}. Five layers of virtual orbitals were optimized in a
series of MCDHF-EOL calculations. Single and double excitations from
the valence $5d$ and the $\{4d,5s,5p\}$ core shells, with at most
one core excitation, were included. The orbitals in Layers 1 to 5
were  $\{6s,6p,6d,4f,5g,6h \}$, $\{7s,7p,7d,5f,6g\}$,
$\{8s,8p,8d,6f\}$, $\{9s,9p,9d\}$, and $\{10s,10p,10d\}$,
respectively.  Step 7  is an  RCI calculation including the CSFs of
Step 6 and also those involving single excitations from the
$\{4s,4p\}$ shells (additional core-valence correlation). In Step 8,
core-valence correlation involving the $\{3s,3p,3d\}$ shells is
added in an RCI calculation. Other RCI calculations including
core-core correlation were made, allowing double excitations from
the $\{5s,5p\}$ shells to Layer 1 (Step 9) and to Layers 1 and 2
(Step 10), in addition to the Step 6 CSFs. In Step 11, double
excitations from the $\{4d,5s,5p\}$ shells to Layer 1 were allowed,
in addition to the Step 6 CSFs. Step 12 is an RCI calculation with a
set of CSFs that is the union of the sets used for Steps 10 and 11.
Step 13 is an RCI calculation that uses the union of the sets of
CSFs used for Steps 7, 10, and 11.  In Step 14, the $\{3s,3p,3d\}$
core-valence contribution (taken as the difference between the
results of Step 8 and Step 7) is added to the results of Step 13. It
was not feasible to include all of the CSFs of Step 13 and Step 8 in
a single RCI calculation.

Table \ref{bacomp} compares the final results with experiment and
with other calculations. The present results are generally in good
agreement with other calculations and with the available
experimental data, with the exception of $A_{5/2}$.  As with
$^{87}$Sr$^+$, the correlation correction to $A_{5/2}$ is so large
as to change its sign relative to the DHF value.  The difference
between the present calculation and the experimental value is 21.4
MHz, which is 31\% of the total correlation correction.

\subsection{Yb$^+$}

The results of the calculations for $^{171}$Yb$^+$ and
$^{173}$Yb$^+$ are given in Table \ref{ybcalc}. The electronic
structures of the lowest-energy states of Yb$^+$ differ from those
of Ba$^+$ in having  fully filled $4f$ shells.  Since the $4f$
electrons are easily excited, correlation effects are expected to be
large. Five layers of virtual orbitals were optimized in a series of
MCDHF-EOL calculations. Single and double excitations from the
valence $5d$ and the $\{4f,5s,5p\}$ core shells, with at most one
core excitation, were included. The orbitals in Layers 1 to 5 were
$\{6s,6p,6d,5f,5g,6h \}$, $\{7s,7p,7d,6f,6g\}$, $\{8s,8p,8d,7f\}$,
$\{9s,9p,9d\}$, and $\{10s,10p,10d\}$, respectively.  Step 7  is an
RCI calculation including the CSFs of Step 6 and also those
involving single excitations from the $\{4s,4p,4d\}$ shells. In Step
8, core-valence correlation involving the $\{3s,3p,3d\}$ shells is
added in an RCI calculation. Step 9 is an RCI calculation allowing
double excitations from the $4f$ shell to Layer 1$^\prime$ (Layer 1
without $6h$), in addition to the Step 6 CSFs. In Step 10, double
excitations from the $\{4f,5s,5p\}$ shells to Layer 1$^\prime$ were
allowed, in addition to the Step 6 CSFs. Step 11 is an RCI
calculation with a set of CSFs that is the union of the sets used
for Steps 7 and 10. In Step 12, the $\{3s,3p,3d\}$ core-valence
contribution (taken as the difference between the results of Step 8
and Step 7) is added to the results of Step 11.

Table \ref{ybcomp} compares the final results for $^{171}$Yb$^+$
with experiment.  It appears that there are no relevant experimental
results for $^{173}$Yb$^+$ nor any comparable calculations for
either isotope.  An experimental value for of $A_{3/2}$ of
$^{171}$Yb$^+$ has been published, but without an estimate of the
uncertainty \cite{engelke96}.  The uncertainty listed in Table
\ref{ybcomp} is based on a private communication \cite{tamm05}.  The
calculated and measured values agree to within this uncertainty. The
sign of the calculated value of $A_{5/2}$ for $^{171}$Yb$^+$ is
correct (unlike the cases for Sr$^+$ and Ba$^+$), but its magnitude
differs from the calculated value by about a factor of 5. The
difference between the present calculation and the experimental
value is 51.1 MHz, which is 28\% of the total correlation
correction. The present calculation of $\mathit{\Theta}_{3/2}$
agrees with the experimental value to within the experimental
uncertainty of 5\%.

\subsection{Hg$^+$}

The results of the calculations for $^{199}$Hg$^+$ and
$^{201}$Hg$^+$ $5d^9 6s^2\;^2 D_{3/2,5/2}$ states are given in Table
\ref{hgcalc}. Hg$^+$ differs from the other ions considered here in
having a more complex electronic configuration.  This necessitated
carefully limiting the CSF expansions to keep the total number of
CSFs per $J$ state below about 45\;000. Four layers of virtual
orbitals were optimized in a series of MCDHF-EOL calculations.
Single and double excitations from the valence $6s$ and the
$\{5s,5p,5d\}$ core shells, with at most one core excitation, were
included. The orbitals in Layers 1 to 4 were  $\{7s,6p,6d,5f,5g,6h
\}$, $\{8s,7p,7d,6f,6g\}$, $\{9s,8p,8d\}$, and $\{10s,9p\}$,
respectively.  The change in the $A$ and $B$ factors upon adding
Layer 4 was on the order of 1\%.  The change in the quadrupole
moments was less than 0.2\%.  In order to limit the numbers of CSFs,
the orbitals of Layer 4 were not used in the RCI calculations.

Steps 6--10 are RCI calculations including the CSFs of Step 4 and
also those involving single excitations from each of the $4f$, $4d$,
$4p$, $4s$, and $3d$ core shells individually (additional
core-valence correlation). The core-valence contributions to the
hyperfine constants are on the order of 1--2\% per shell for the
$n=4$ shells, but less for the $3d$ shell.  The corresponding
contributions to the quadrupole moments are small, less than 0.2\%
per shell. Step 11 is an RCI calculation allowing double ($d$)
excitations from the $\{5d,6s\}$ shells to Layer 1$^\prime$
($\{7s,6p,6d,5f\}$), in addition to the Step 4 CSFs. In Step 12,
double and triple ($dt$) excitations from the $\{5d,6s\}$ shells to
Layer 1$^\prime$ were allowed, in addition to the Step 4 CSFs.
Significant changes in both the hyperfine constants and the
quadrupole moments were noted in both Step 11 and Step 12. Step 13
is an RCI calculation with a set of CSFs that is the union of the
set used for Step 12 and the set obtained by allowing double
excitations from the $\{5s,5p,5d\}$ shells to Layer 1$^\prime$.  In
Step 14, the core-valence contributions from the
$\{3d,4s,4p,4d,4f\}$ shells calculated in separate RCI calculations
(Steps 6--10) are added to the results of Step 13.

The validity of adding core-valence contributions from separate RCI
calculations was verified by comparing the results for pairs of core
shells considered together and separately.  For example, the $4s$
core-valence RCI calculation (Step 9) changes $A_{5/2}$ of
$^{199}$Hg$^+$ by +19.6 MHz compared to the Step 4 MCDHF result. The
$4p$ core-valence RCI calculation (Step 8) changes it by +11.1 MHz.
An RCI calculation in which the $4s$ and $4p$ core-valence
contributions were both included resulted in a change of +31.1 MHz,
compared to +30.7 MHz for the sum of the $4s$ and $4p$ contributions
calculated separately.  It was not feasible to include all of the
core-valence contributions in a single RCI calculation.

Table \ref{hgcomp} compares the final results for $^{199}$Hg$^+$ and
$^{201}$Hg$^+$ with experiment and with other calculations.  The
present result for $A_{5/2}$ of $^{199}$Hg$^+$ agrees within 2.4\%
with the experimental result.   No experimental values for the $B$
factors of $^{201}$Hg$^+$ are available for comparison. Observations
of the hyperfine structure by classical optical spectroscopy
\cite{mrozowski40,loebich62} are not precise enough for this
purpose. The calculations of Brage \textit{et al.} \cite{brage99}
were carried out by the  MCDHF and RCI methods with a set of CSFs
more limited than that for the present calculation.   The
experimental value for $\mathit{\Theta}_{5/2}$ is 26\% smaller in
magnitude than the DHF value, so the correlation contribution to the
quadrupole moment is greater than for the other ions studied here.
The present result for $\mathit{\Theta}_{5/2}$ disagrees with the
experimental value by about 10.5\%, which is about 3 times the
experimental uncertainty. The disagreement is about 30\% of the
total correlation contribution.

\subsection{Au}

The results of the calculations for the $^{197}$Au $5d^9 6s^2\;^2
D_{3/2,5/2}$ states are given in Table \ref{aucalc}. The steps in
the calculation are the same as for Hg$^+$ (Table \ref{hgcalc}).
Table \ref{aucomp} compares the final results with experiment. The
calculated values of $A_{3/2}$ and $A_{5/2}$ agree with the
experimental values to within 3\%. The calculated values of
$B_{3/2}$ and $B_{5/2}$ differ from experiment by about 8\%.
However, this comparison depends on the value assumed for the
nuclear quadrupole moment $Q$($^{197}$Au).

The current status of knowledge of $Q$($^{197}$Au) has been
summarized by Schwerdtfeger \textit{et al.} \cite{schwerdtfeger05}.
The currently accepted value listed by Pyykk\"{o} \cite{pyykko01} is
+0.547(16) b (1 b = 10$^{-28}$ m$^2$) and is based on muonic
hyperfine measurements.  A value of +0.594(10) b was derived by
Blachman \textit{et al.} \cite{blachman67} based on the experimental
$B$ factors of the Au $5d^9 6s^2\;^2 D_{3/2,5/2}$ states
\cite{childs66,blachman67}, i.e., the same states studied in the
present work.  However, Blachman \textit{et al.} did not calculate
the atomic electric field gradients from \textit{ab initio} theory,
as in the present work, but inferred them from the experimental $A$
factors. This method is of uncertain accuracy and does not include
the Sternheimer antishielding correction, which is included in the
present calculation. Schwerdtfeger \textit{et al.}
\cite{schwerdtfeger05} obtained $Q$($^{197}$Au) = +0.60 b from
measured M\"{o}ssbauer electric quadrupole splittings in a large
number of gold compounds combined with solid-state
density-functional calculations.  They obtained $Q$($^{197}$Au) =
+0.64 b from the measured electric quadrupole coupling constant in
(CO)AuF (i.e., the weakly bound complex of a CO molecule and a AuF
molecule), together with a relativistic coupled-cluster calculation
of the electronic structure of the complex.

Since the muonic and the other determinations of $Q$($^{197}$Au)
appear to be discrepant, it is of interest to derive a value based
on the present calculations.  These calculations imply values of
$Q$($^{197}$Au) of +0.5918 b and +0.5816 b, based on the
experimental values of $B_{3/2}$ and $B_{5/2}$, respectively.  These
values agree to within about 2\%.  Assigning an uncertainty to the
value of $Q$ derived in this way is difficult.  One way is to make
use of the fact that the relative calculational errors for the $A$
and $B$ factors are similar, since they both depend to first order
on matrix elements of $1/r^3$ for the $5d$ electrons. (This method
of estimating the relative errors would not hold for cases where
there are large cancelations, as for $A_{5/2}$ in Ca$^+$, Sr$^+$,
Ba$^+$, and Yb$^+$.) Based on the fact that the calculated values of
$A_{3/2}$ and $A_{5/2}$ for $^{197}$Au and $A_{5/2}$ for
$^{199}$Hg$^+$ agree with the experimental values to within 3\% or
better, 5\% is a reasonable estimate for the error in the estimate
of $Q$.  The new estimate, based on the average of the values
derived from $B_{3/2}$ and $B_{5/2}$ of $^{197}$Au, is
$Q$($^{197}$Au) = +0.587(29) b. The error bars of the present
measurement overlap those of the muonic measurement.

\section{Conclusions}

The main object of this study was to calculate the atomic quadrupole
moments of the metastable $^2D_{3/2,5/2}$ states of several ions and
atoms to an uncertainty better than those of the simple estimates
obtained from Hartree-Fock or Dirac-Hartree-Fock calculations [e.g.,
Eqs.(\ref{singleconf1a}) and (\ref{singleconf1b})].  This is
apparently the first use of MCDHF and RCI methods for this purpose.
For Ca$^+$, Ba$^+$, and Au, there are no experimental or other
theoretical values for comparison.  For $\mathit{\Theta}_{5/2}$ of
Sr$^+$, the experimental determination has an uncertainty of 11.5\%,
so it does not provide a precise test of the calculation. However, a
recent relativistic coupled-cluster calculation \cite{sur05} agrees
with the present calculation to within 4\%. An experimental
determination of $\mathit{\Theta}_{3/2}$ of Yb$^+$ agrees with the
present calculation to within the experimental uncertainty of 5\%.
The experimental determination of $\mathit{\Theta}_{5/2}$ for Hg$^+$
differs from the present calculation by 10.5\%. In summary, the
method used in this work appears to be capable of calculating the
atomic quadrupole moments to about 5\% or better for configurations
consisting of a single $nd$ electron outside a set of closed shells,
while the error appears to be about 10\% for the more complex $5d^9
6s^2\;^2 D_{3/2,5/2}$ states.

The second object was to calculate the hyperfine constants of the
same states.  For most of the cases where there is experimental
data, the agreement is within a few percent.  For the $B$ factors,
some of the discrepancies may be due to errors in the nuclear
quadrupole moments used.  The exception to the generally good
agreement is for the $A$ factors of the $^2 D_{5/2}$ states of
Sr$^+$, Ba$^+$, and Yb$^+$, where the correlation contributions
exceed 100\% of the Dirac-Hartree-Fock values, leading to a change
in sign of the constants, relative to the DHF values. The present
calculations are in error by about 30\% of the total correlation
contribution. The source of the error is not understood.  It may be
related to limitations on the form of the CSFs included in the
calculations or to the particular strategy used for the optimization
of the orbitals.  Apparently, many-body perturbation theory or
coupled-cluster theory can give better results for the $A$ factors
of these states, although this has not yet been demonstrated for
Yb$^+$.

The present methods  give good results for the hyperfine constants
of the $5d^9 6s^2\;^2 D_{3/2,5/2}$ states of Hg$^+$ and Au.  The $A$
factors agree with experiment to about 3\%.  The $B$ factors
calculated for Au disagree by about 8\%, but this may  be due to an
error in the currently accepted value of the nuclear quadrupole
moment. The present calculations are  the most accurate \emph{ab
inito} calculations for the hyperfine constants of these states.
Apparently, many-body perturbation theory or coupled-cluster theory
have not yet been applied to these systems.

\begin{acknowledgments}
I thank B. P. Das and colleagues for communicating their theoretical
results and Chr. Tamm for communicating experimental results prior
to publication. I thank Prof. C. Froese Fischer and Prof. G.
Gaigalas for reading the manuscript and providing information about
the \textsc{GRASP} codes. This research was partially supported by
the Office of Naval Research.
\end{acknowledgments}


\end{document}